\documentclass[11pt]{amsart}

\usepackage{amsmath}
\usepackage{amssymb,mathtools,amsthm,latexsym}
\usepackage{mathrsfs}
\usepackage{enumitem}
\usepackage{bm}
\usepackage{url}
\usepackage{xcolor}
\usepackage{pgfplots}
\usepgfplotslibrary{fillbetween}
\pgfplotsset{width=0.55\textwidth}
\numberwithin{equation}{section}

\usepackage[nocompress]{cite}

\calclayout

\usepackage{comment}
\usepackage{soul}

\newtheorem{thm}{Theorem}[section]

\theoremstyle{definition}

\newtheorem*{ack}{Acknowledgements}
\newtheorem{remark}[thm]{Remark}
\newtheorem{prop}[thm]{Proposition}

\def\XXint#1#2#3{{\setbox0=\hbox{$#1{#2#3}{\int}$}
		\vcenter{\hbox{$#2#3$}}\kern-.5\wd0}}

\newcommand{\R}{\ensuremath{\mathbb{R}}}

\newcommand{\defeq}{\mathrel{\mathop:}=}

\newcommand{\ud}{\mathrm{d}}

\newcommand{\inv}{^{-1}}

\title[Causal bubbles]{Causal bubbles in globally hyperbolic spacetimes}
\author{Leonardo Garc\'{i}a-Heveling, Elefterios Soultanis}

\address{Leonardo Garc\'{i}a-Heveling\\
	Department of Mathematics\\
	Radboud University\\
	P.O. Box 9010, Postvak 59\\
	6500 GL Nijmegen\\
	The Netherlands}
\email{\tt l.heveling@math.ru.nl}

\address{Elefterios Soultanis\\
	Department of Mathematics\\
	Radboud University\\
	P.O. Box 9010, Postvak 59\\
	6500 GL Nijmegen\\
	The Netherlands}
\email{\tt elefterios.soultanis@gmail.com}

\begin{document}
	
\maketitle

\begin{abstract}
    We give an example of a spacetime with a continuous metric which is globally hyperbolic and exhibits causal bubbling. The metric moreover splits orthogonally into a timelike and a spacelike part. We discuss our example in the context of energy conditions and the recently introduced synthetic timelike curvature-dimension (TCD) condition. In particular we observe that the TCD-condition does not, by itself, prevent causal bubbling.
\end{abstract}
	
\section{Introduction}

Spacetimes where the Lorentzian metric is merely continuous often appear in mathematical General Relativity as weak solutions of Einstein's Equations \cite{BuLF14,GeTr87,Sbi18}. Chru\'sciel and Grant \cite{ChrGra12} were the first to study their causal structure, discovering a phenomenon called \emph{causal bubbling} (see below). Later, S\"amann \cite{Sae16} studied globally hyperbolic spacetimes with continuous metrics in detail and showed that, just as in the smooth case, global hyperbolicity can be equivalently characterized by any of the following:
\begin{enumerate}[label=(\roman*)]
    \item Non-total imprisonment and compact causal diamonds,
%{\color{blue}    \item Non-total imprisonment and compactness of the space of causal curves, ??}
    \item Existence of a Cauchy hypersurface,
    \item Existence of a Cauchy time function.
\end{enumerate}
In the same paper (see Section 7) S\"amann raises the question whether globally hyperbolic spacetimes can be causally bubbling. In this short note we give an example of a {\bf globally hyperbolic spacetime with continuous metric $-\ud t^2+\rho\ud x^2$ exhibiting causal bubbling}.

We define causal bubbling on a spacetime $(M,g)$ with continuous metric $g$ as follows. Set
\begin{align*}
    \mathcal B^\pm(p)=J^\pm(p)\setminus \overline{I^\pm(p)}
\end{align*}
The set $\mathcal B^\pm(p)$ is called a (future/past) \emph{causal bubble} if it is non-empty.\footnote{This definition corresponds to \emph{external} bubbling in \cite{GKSS20} and is equivalent to the failure of the \emph{push up} property, see \cite[Theorem 2.12]{GKSS20}.} Here $I^\pm(p)$ and $J^\pm(p)$ are the timelike and causal future/past cones defined using \emph{Lipschitz} (or equivalently absolutely continuous) curves. We refer to \cite{GKSS20} for a discussion on timelike and causal cones defined via smooth curves and their relationship with the notion used here. See also \cite{Leo21,Ling20} and \cite[Section 5.1]{kun-sae18} for further analyses on causality with continuous metrics.%\footnote{In particular the authors in \cite{GKSS20} define \emph{interior} and \emph{external} bubbling; our notion corresponds to external bubbling.} 

Our example stands out from previously known examples of causally bubbling spacetimes for two reasons.
\begin{enumerate}[label=(\alph*)]
    \item It is manifestly globally hyperbolic,
    \item The metric splits orthogonally into a timelike and a spacelike part.
\end{enumerate}
The second statement (b) would follow automatically from (a) if the metric were at least $\mathcal{C}^2$ \cite{BerSan05}, but not if it is merely continuous. In particular, while examples 3.1 and 3.2 in \cite{GKSS20} (which exhibit \emph{internal} bubbling) likely are globally hyperbolic, it is not clear whether the metric splits orthogonally as in (b). Whether Example 1.11 in \cite{ChrGra12} is globally hyperbolic is less clear, but it is known to be strongly causal (see \cite[Section 5.1]{kun-sae18}).

\section{The example}

Consider the $(1+1)$-dimensional spacetime $\R^2$ equipped with the continuous Lorentzian metric
\begin{align} \label{eq:metric}
g :=-\ud t^2+\rho(t,x)\ud x^2,\quad \rho(t,x):=1+\sqrt{(t-|x|)_+}.
\end{align}
With the natural choice of time orientation, $t$ is a time function. Since the lightcones of the metric \eqref{eq:metric} are narrower than those of the Minkowski metric, $t$ is a Cauchy time function, and hence the spacetime is globally hyperbolic.

To see that the causal future of the origin contains a causal bubble, we begin by considering the ODE for the null curves $\gamma(s)=(\alpha(s),\beta(s))$:

\begin{align*}
0=-\alpha'(s)^2+\left(1+\sqrt{(\alpha(s)-|\beta(s)|)_+}\right)\beta'(s)^2.
\end{align*}

A null curve $\gamma$ starting at the origin, parametrized as $(\alpha(s),s)$ thus satisfies $\alpha'(s)^2=1+\sqrt{(\alpha(s)-s)_+}$ or, by denoting $y(s)=\alpha(s)-s\ge 0$,
\begin{align}\label{eq:IVP}
y'(s)+1=\sqrt{1+\sqrt{y(s)}},\quad y(0)=0.
\end{align}
In addition to the trivial solution $y\equiv 0$, the initial value problem \eqref{eq:IVP} also admits another solution, expressed in implicit form as
\begin{align*}
s=\frac 43\left[\left(1+\sqrt{y(s)}\right)^{3/2}-1\right]+2\sqrt{y(s)}.
\end{align*}
Indeed, differentiating both sides yields
\begin{align*}
	1 &=2\frac{\ud}{\ud s}\sqrt{y(s)}(1+\sqrt{y(s)})^{1/2}+2\frac{\ud}{\ud s}\sqrt{y(s)}\\
	 &=y'(s)\frac{\sqrt{1+\sqrt{y(s)}}+1}{\sqrt{y(s)}}=\frac{y'(s)}{\sqrt{1+\sqrt{y(s)}}-1}.
\end{align*}
Denoting
\begin{align*}
	f(y):=\frac 43 \left[\left(1+\sqrt y\right)^{3/2}-1\right]+2\sqrt y, 
\end{align*}
we conclude that $\gamma=(s+f\inv(s),s)$ is a null curve as well as the straight line given by $\gamma=(s,s)$. In fact there is a 1-parameter family of null curves starting at $0$, given by
\begin{align} \label{eq:nullgeodesics}
    \gamma_u(s) := \begin{cases} (s,s) &\textrm{for } 0 \leq s < u, \\
    (s+f\inv(s-u),s) &\textrm{for } s \geq u,
    \end{cases}
\end{align}
where $u \in [0,\infty]$. One can check that the curves $\gamma_u$ are %not $\mathcal C^2$ but instead $\mathcal C^{1,1}$ regular. 
smooth for all parameter values except $s=u$, where they are only $\mathcal C^{1,1}$ regular. Note also that on smooth, two-dimensional spacetimes, every null curve is a null geodesic. Hence, for $u<\infty$, $\gamma_u \vert_{(u,\infty)}$ is a null geodesic, as it is contained in the region where the metric is smooth. At $s=u$, however, $\gamma_u$ is not locally length maximizing.

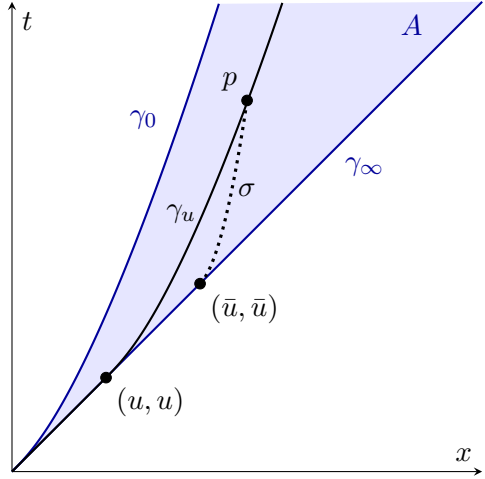
\begin{figure}
\begin{center}
   \begin{tikzpicture}
   \tikzset{declare function={a=100;}}
    \begin{axis}[xmin=0,xmax=a,
                 ymin=0,ymax={a},
                 axis equal image,
                 axis on top=true,
                 axis x line=middle,
                 axis y line=middle,
                 xlabel={$x$},
                 ylabel={$t$},
                 xtick=\empty,
                 ytick=\empty,
                 clip=false]
     \addplot [domain=0:44, samples=100, smooth, name path=A, thick,blue!60!black] {x+((0.16667*(1.4422*(27*x^2+1.7321*sqrt(243*x^4+3024*x^3+34776*x^2+208512*x+439024)+72*x+12)^(2/3)-24.961*x-158.09))/((27*x^2+1.7321*sqrt(243*x^4+3024*x^3+34776*x^2+208512*x+439024)+72*x+12)^(1/3)))^2};
     \draw[name path = B, thick,
     blue!60!black
     ] (axis cs:0,0)--(axis cs:a,a);
     \draw[thick,
     ] (axis cs:0,0)--(axis cs:20,20);
     \addplot [domain=0:37.5, samples=50, thick, name path=C] (x+20,{x+20+((0.16667*(1.4422*(27*x^2+1.7321*sqrt(243*x^4+3024*x^3+34776*x^2+208512*x+439024)+72*x+12)^(2/3)-24.961*x-158.09))/((27*x^2+1.7321*sqrt(243*x^4+3024*x^3+34776*x^2+208512*x+439024)+72*x+12)^(1/3)))^2});
     \addplot [domain=0:10, samples=40, very thick, smooth, dotted] (x+40,40+x+0.29*x*x);
     \addplot[blue!10] fill between[of = A and B];
     \filldraw[black] (axis cs: 50,79) circle (2pt) node[anchor=south east] {$p$};
     \filldraw[black] (axis cs: 40,40) circle (2pt) node[anchor=north west] {$(\bar{u},\bar{u})$};
     \filldraw[black] (axis cs: 20,20) circle (2pt) node[anchor=north west] {$\textcolor{black}{(u,u)}$};
     \draw (axis cs: 85, 95) node {\textcolor{blue!60!black}{$A$}};
     \draw (axis cs: 28,75) node {\textcolor{blue!60!black}{$\gamma_0$}};
     \draw (axis cs: 75, 65) node {\textcolor{blue!60!black}{$\gamma_\infty$}};
     \draw (axis cs: 36, 55) node {\textcolor{black}{$\gamma_u$}};
     \draw (axis cs: 50, 60) node {$\sigma$};
     \end{axis}
   \end{tikzpicture}
   
  \caption{The causal bubble $A$ as in Proposition \ref{prop:bubble}.}
  \label{fig1}
\end{center}
\end{figure}

\begin{prop}\label{prop:bubble}
	The non-empty open set $A:=\{(t,x): x>0,\ 0<t-x< f\inv(x) \}\subset \R^2$ consists of points in $J^+(0)\setminus \overline{I^+(0)}$.
\end{prop}
Consequently $\R^2$ equipped with the globally hyperbolic metric \eqref{eq:metric} contains causal bubbles.
\begin{proof}
The inclusion $A\subset J^+(0)$ is clear. Let $p \in A$. We first show that $p \not\in I^+(0)$. Since the set $A$ is foliated by the null curves \eqref{eq:nullgeodesics}, there is a unique $u \in (0,\infty)$ such that $\gamma_u$ passes through $p$. Because the metric is smooth outside of $t = \pm x$, $\gamma_u$ is a null geodesic generating the boundary of $I^-(p)$, at least from $(u,u)$ until $p$. This means that any past-directed timelike curve $\sigma$ starting at $p$ must intersect the diagonal at some point $(\bar{u},\bar{u})$ with $\bar{u} \geq u > 0$ (see Figure \ref{fig1}). It follows that $\sigma$ cannot reach $0$; indeed, following the diagonal would introduce a null piece, while leaving the diagonal violates the causality of the curve (the metric \eqref{eq:metric} has narrower lightcones than those of the Minkowski metric). We conclude that $0 \not\in I^-(p)$, (equivalently, $p \not\in I^+(0)$) and hence $A \cap I^+(0) = \emptyset$. But since $A$ is open, it cannot contain any boundary points of $I^+(0)$ either, hence also $A \cap \overline{I^+(0)} = \emptyset$, concluding the proof.
\end{proof}

\section{Discussion}
\subsection{Strong energy condition}
As in previously known examples \cite{ChrGra12,Leo21,GKSS20,Ling20} causal bubbling arises from the branching of null geodesics. Chru\'sciel and Grant noted \cite[Rem.\ 1.19]{ChrGra12} that, in the Riemannian case, branching is associated with curvature being unbounded from below. Indeed, in our example (in $\{ t\ne |x|\}$) the Ricci scalar, given by
%One might speculate that the same is true in the Lorentzian case. Indeed, in our example, on the region where the metric is smooth, the Ricci scalar is given by
\begin{equation*} %\label{eq:curvature}
    R = \begin{cases}
    -\frac{\rho + (\rho - 1)^2}{4 \rho^2 (\rho - 1)^3} & \textrm{ if } t > \vert x \vert \\
    0 & \textrm{ if } t < \vert x \vert,
    \end{cases}
\end{equation*}
diverges to $-\infty$ as $t \searrow \vert x \vert$. Note however that in dimension $1+1$ the Ricci \emph{tensor} is given by $\operatorname{Ric} = \frac{1}{2} R g$ and thus
\begin{equation}\label{eq:ricci_non_negative}
{\rm Ric}(v,v)\ge 0\quad \textrm{ for all causal vectors $v$ in $\{ t\ne |x|\}$}.
\end{equation}
In other words the strong energy condition is satisfied. Physically speaking causal bubbling appears to be a consequence of the presence of infinite (but positive) effective energy density (see \cite{KoKo20} for an in-depth discussion of energy conditions).

\subsection{Synthetic curvature bounds}

Recently, a \emph{synthetic} notion of \emph{timelike curvature dimension (TCD) bounds}  on (non-smooth) \emph{Lorentzian pre-length spaces} has been put forth using optimal transport \cite{McC20,CM20}, in analogy with the very successful metric theory \cite{stu06a,stu06b,lot09,ags14b}. The \emph{entropic convexity} condition defining TCD$(K,N)$-spaces asks that the \emph{R\'enyi entropy}
\[
{\rm Ent}(f\operatorname{vol})\defeq \int f\log f\ud\operatorname{vol}
\]
is $(K,N)$-convex along timelike geodesics in a space of probability measures. We refer to \cite[Definition 2.8]{kun-sae18} and \cite[Definition 2.17]{Braun22} (see also \cite{CM20,CM22}) for the definitions and properties of Lorentzian pre-length spaces and $(K,N)$-convexity, respectively. On smooth spacetimes, such convexity properties characterize the strong energy condition, cf. \cite{McC20}.

%Following the success of \emph{synthetic} curvature bounds in (non-smooth) metric spaces [CITATIONS], a definition of \emph{timelike curvature bounds} on \emph{Lorentzian length spaces} in a similar spirit has been put forth in the last few years in [CITE], framed in terms of convexity properties of the Renyi entropy

%curvature bounds in terms of convexity properties of an entropy functional (the relative Renyi entropy) along geodesics in a space of probability measures. On smooth spacetimes, such convexity properties characterize the strong energy condition, cf. [McCann]. We refer to [CITE] and [CITE] for the definition of Lorentzian length spaces and the TCD-condition, respectively.

Despite satisfying \eqref{eq:ricci_non_negative} the R\'enyi entropy associated to the volume measure of the metric \eqref{eq:metric} is not $(K,N)$-convex along Lorentz-Wasserstein geodesics, i.e. {\bf the metric \eqref{eq:metric} does not satisfy the entropic convexity condition for any $(K,N)$}.\footnote{While Lorentzian pre-length spaces are required (by definition) to have the push up property the entropic convexity condition makes sense regardless of the validity of push up.} This follows from the fact that $t\mapsto -\log\rho(t,x)$ is not $(K,N)$-convex for any $x\in \R$, cf. \cite{Sou22}.

However, restricted to a suitable subset, our example demonstrates that the TCD-condition does not prevent causal bubbling. Indeed, the closed set $Y\defeq \{t\ge |x|\}\subset \R^2$ with the restriction of the metric \eqref{eq:metric} satisfies the (weak) entropic $(0,2)$-convexity condition but contains causal bubbles. Notice that $(Y,g)$ is obtained as the uniform pointwise limit of the sequence of smooth metrics 
\begin{align}\label{eq:seq}
g_j\defeq-\ud t^2+\rho_j(t,x)\ud x^2,\quad \rho_j(t,x)=\rho(t+1/j,x)
\end{align}
on $Y$, which all satisfy the strong energy condition \eqref{eq:ricci_non_negative} everywhere on $Y$ and are thus TCD$(0,2)$-spaces.\footnote{Note that, while the spaces $(Y,g_j)$ are Lorentzian \emph{pre-}length spaces, every point on $\{t=|x|\}\subset Y$ has empty chronological past and thus they fail to be Lorentzian length spaces, cf. \cite[Definition 3.22]{kun-sae18}.}

In closing we point out that, in the metric setting, branching of geodesics is excluded by the \emph{Riemannian curvature-dimension} (RCD) condition but not by the (weak) CD-condition, see \cite{deng21} and \cite{mag22}, respectively.

\begin{remark}
Note that all the conclusions in this note, including those about curvature, are valid in higher dimensions as seen by considering metrics $g=-\ud t^2+\rho(t,|\bm x|)\ud \bm x^2$ in $\R^{n+1}$.
\end{remark}

 \begin{ack}
We thank Annegret Burtscher, Eric Ling and Clemens S\"amann for interesting discussions and comments.
 \end{ack}

\bibliographystyle{abbrv}
\bibliography{abib}
	
\end{document}